\documentclass{article}
\usepackage{amsmath, amssymb, amsfonts}
\title{Bjorken expansion in the isotropic Kasner spacetime}
\author{Hristu Culetu, \\Ovidius University, Dept.of Physics, \\B-dul Mamaia 124, 8700 Constanta, Romania, \\e-mail : hculetu@yahoo.com}

\begin{document}
\numberwithin{equation}{section}
\pagenumbering{arabic}
\maketitle
\newcommand{\fv}{\boldsymbol{f}}
\newcommand{\tv}{\boldsymbol{t}}
\newcommand{\gv}{\boldsymbol{g}}
\newcommand{\OV}{\boldsymbol{O}}
\newcommand{\wv}{\boldsymbol{w}}
\newcommand{\WV}{\boldsymbol{W}}
\newcommand{\NV}{\boldsymbol{N}}
\newcommand{\hv}{\boldsymbol{h}}
\newcommand{\yv}{\boldsymbol{y}}
\newcommand{\RE}{\textrm{Re}}
\newcommand{\IM}{\textrm{Im}}
\newcommand{\rot}{\textrm{rot}}
\newcommand{\dv}{\boldsymbol{d}}
\newcommand{\grad}{\textrm{grad}}
\newcommand{\Tr}{\textrm{Tr}}
\newcommand{\ua}{\uparrow}
\newcommand{\da}{\downarrow}
\newcommand{\ct}{\textrm{const}}
\newcommand{\xv}{\boldsymbol{x}}
\newcommand{\mv}{\boldsymbol{m}}
\newcommand{\rv}{\boldsymbol{r}}
\newcommand{\kv}{\boldsymbol{k}}
\newcommand{\VE}{\boldsymbol{V}}
\newcommand{\sv}{\boldsymbol{s}}
\newcommand{\RV}{\boldsymbol{R}}
\newcommand{\pv}{\boldsymbol{p}}
\newcommand{\PV}{\boldsymbol{P}}
\newcommand{\EV}{\boldsymbol{E}}
\newcommand{\DV}{\boldsymbol{D}}
\newcommand{\BV}{\boldsymbol{B}}
\newcommand{\HV}{\boldsymbol{H}}
\newcommand{\MV}{\boldsymbol{M}}
\newcommand{\be}{\begin{equation}}
\newcommand{\ee}{\end{equation}}
\newcommand{\ba}{\begin{eqnarray}}
\newcommand{\ea}{\end{eqnarray}}
\newcommand{\bq}{\begin{eqnarray*}}
\newcommand{\eq}{\end{eqnarray*}}
\newcommand{\pa}{\partial}
\newcommand{\f}{\frac}
\newcommand{\FV}{\boldsymbol{F}}
\newcommand{\ve}{\boldsymbol{v}}
\newcommand{\AV}{\boldsymbol{A}}
\newcommand{\jv}{\boldsymbol{j}}
\newcommand{\LV}{\boldsymbol{L}}
\newcommand{\SV}{\boldsymbol{S}}
\newcommand{\av}{\boldsymbol{a}}
\newcommand{\qv}{\boldsymbol{q}}
\newcommand{\QV}{\boldsymbol{Q}}
\newcommand{\ev}{\boldsymbol{e}}
\newcommand{\uv}{\boldsymbol{u}}
\newcommand{\KV}{\boldsymbol{K}}
\newcommand{\ro}{\boldsymbol{\rho}}
\newcommand{\si}{\boldsymbol{\sigma}}
\newcommand{\thv}{\boldsymbol{\theta}}
\newcommand{\bv}{\boldsymbol{b}}
\newcommand{\JV}{\boldsymbol{J}}
\newcommand{\nv}{\boldsymbol{n}}
\newcommand{\lv}{\boldsymbol{l}}
\newcommand{\om}{\boldsymbol{\omega}}
\newcommand{\Om}{\boldsymbol{\Omega}}
\newcommand{\Piv}{\boldsymbol{\Pi}}
\newcommand{\UV}{\boldsymbol{U}}
\newcommand{\iv}{\boldsymbol{i}}
\newcommand{\nuv}{\boldsymbol{\nu}}
\newcommand{\muv}{\boldsymbol{\mu}}
\newcommand{\lm}{\boldsymbol{\lambda}}
\newcommand{\Lm}{\boldsymbol{\Lambda}}
\newcommand{\opsi}{\overline{\psi}}
\renewcommand{\tan}{\textrm{tg}}
\renewcommand{\cot}{\textrm{ctg}}
\renewcommand{\sinh}{\textrm{sh}}
\renewcommand{\cosh}{\textrm{ch}}
\renewcommand{\tanh}{\textrm{th}}
\renewcommand{\coth}{\textrm{cth}}

\begin{abstract}
An isotropic expansion for the QGP is proposed in curved Kasner spacetime for an experimental configuration with three dimensional set of beams. The fluid of relativistic particles has no shear viscosity but the nonzero bulk viscosity $\zeta$ is time dependent and its value could explain the enormous entropy per baryon of our Universe. In addition, $\zeta$ equals the bulk viscosity of the anisotropic compressible fluid conjectured for the interior of a black hole.
\end{abstract}

 Keywords : isotropic expansion ; bulk viscosity ; Kasner spacetime.\\

\textbf{Introduction}

 One of the most important lessons one may learn from experiments on heavy - ion collisions at high energies (RHIC accelerator, Brookhaven) is that the fluid hydrodynamics are relevant for understanding the dynamics of the phenomenon \cite{SH, SO}. Most hydrodynamic simulations which describe the Bjorken elliptic flow are consistent with an almost perfect fluid behaviour ( a small ratio $\eta/s$ - shear viscosity/entropy density). The boost - invariant quark gluon plasma (QGP) - the deconfined phase of QCD, characterized by collective degree of freedom, behaves as a relativistic expanding anisotropic fluid  where particle and spatial rapidities are identical \cite{MH}. 
 
 Since a perturbative description leads in general to a high ratio $\eta/s$, which is not in agreement with experiments, a strong coupling regime of QCD is more appropriate for the RHIC study. The new tool of the gauge/gravity correspondence (AdS/CFT duality) offers a new viewpoint on the phenomenon (the 5 - dimensional metric of the AdS space has the 4 - dimensional Mionkowski space as boundary).
 
 K. Kajantie et al. \cite{KK} studied the AdS/CFT thermodynamics of the spatially isotropic configuration of the Bjorken flow in d - dimensional Minkowski space, with $d \geq 3$. The bulk solution is a nonstatic Schwarzschild - AdS black hole while the boundary matter is an isotropic expanding perfect fluid from a pointlike explosion. Due to the spatial isotropy, the fluid flow has a vanishing shear tensor. The shear viscosity does not therefore contribute to the stress tensor. In addition, the prediction $\eta/s = \hbar/2 \pi k_{B}$ cannot be checked, where $k_{B}$ is the Boltzmann constant. However, the nonstatic character of the fluid flow leads to a nonzero scalar expansion and, hence, a bulk viscosity term will appear in the energy - momentum tensor \cite{KK}. 
 
 In \cite{BO} Brevik and Odintsov applied the isotropic expansion for a cosmic fluid endowed with bulk viscosity, generalizing the Cardy - Verlinde entropy formula. Later, Bravik et al. \cite{BGG} showed how the bulk viscosity $\zeta$ can influence the future singularity for a FLRW universe (The Big Rip, when one or more of the physical quantities go to infinity in a finite time in the future). They introduced the Casimir effect in Cosmology through the Casimir energy $E_{C}$, inversely proportional to the scale factor $a(t)$. They also found that, at late times when $a(t) \rightarrow \infty$, the Casimir influence fades away.
 
 Tawfik et al. \cite{TWMH} assumed that $\zeta$ plays an essential role in the early Universe, especially in the region close to the QCD critical temperature $T_{c}$. Recent lattice QCD calculations at high temperature help to determine the bulk viscosity of the Quark Gluon Plasma (QGP). The authors also studied the evolution equation of the Hubble parameter $H = \dot{a}(t)/a(t)$ for a FLRW universe filled with a viscous QGP, where $a(t)$ is the scale factor. The validity of the model is restricted to the QCD era. Tawfik and Harko \cite{TH} studied the quark - hadron phase transition by taking into account the effect of the bulk viscosity when an out-of-equilibrium nucleation of hadron bubbles in the QGP surrounding should take place. The scale of the cosmological quark - gluon phase transition is given by the Hubble radius $R_{H} = m_{P}/T_{c}^{2}$, where the mass inside the Hubble volume is about one solar mass. The authors conjectured that the above phase transition might take place in the RHIC process.
 
 A cosmic fluid with $\zeta \neq 0$ is also able to produce a Little Rip Cosmology \cite{BENO} (when singularity occurs at the infinite future) which is considered to be a viable alternative to $\Lambda CDM$ cosmology (see also \cite{MG}). 
 
 We shall apply in this paper Brevik's and Pettersen's ideas \cite{BP} on viscous cosmology in the Kasner metric, for the RHIC (the similarities between the state of the very early stages of our Universe and a system of ultrarelativistic particles in collision are well known ). We analyse a nonflat isotropic Kasner spacetime created by a stress tensor from the r.h.s. of Einstein's equations corresponding to a fluid with nonzero time dependent bulk viscosity. 
 
 Because of the lack of the shear viscosity $\eta$, the model is appropriate for an experimental device with three beam axes ( a 3 - dimensional expansion \cite{SS} for the RHIC fireball). Even though the authors of \cite{SS} treated an extension of the one dimensional Bjorken expansion using anisotropic Kasner spacetime, we think the isotropic version is suitable for an experiment with three beam axis or a spherically symmetric system of beams.\\
 Throughout the paper we take $c = G = k_{B} = 1$. \\
 
 \textbf{The nonflat Kasner metric}
 
 Let us consider the well known form of the Kasner line element \cite{BP}
 \begin{equation}
 ds^{2} = - dt^{2} + t^{2a} dx^{2} + t^{2b} dy^{2} + t^{2c} dz^{2}, 
 \label{1}
 \end{equation}
 where $a, b ~and~ c$ are constants and (1) is a solution of the vacuum Einstein equations if 
 \begin{equation}
 a + b + c = 1,~~~~a^{2} + b^{2} + c^{2} = 1.
 \label{2}
 \end{equation}
 In this paper we will be concerned with a viscous fluid which is consider as a source of curvature (its stress tensor $T_{\mu \nu}$ will play the role of a source in the r.h.s. of Einstein 's equations). Therefore, Eqs. (2) will no longer be valid.
 
 The equations of gravity will be written as 
 \begin{equation}
 R_{\mu \nu} - \frac{1}{2} g_{\mu \nu} R = 8 \pi T_{\mu \nu} .
 \label{3}
 \end{equation}
 In what follows we use the positive signature (-, +, +, +) and $R_{\mu \nu} \equiv R^{\alpha}_{\mu \alpha \nu}$. The Greek indices run from 0 to 3.\\
 The nonzero components of the Ricci tensor are given by 
 \begin{equation}
 R_{tt} = [ a + b + c - (a^{2} + b^{2} + c^{2})] t^{-2},~~~~R_{xx} = a (a + b + c - 1) t^{2a - 2} ,
 \label{4}
 \end{equation}
 with similar expressions for $R_{yy}~ and~ R_{zz}$ as for $R_{xx}$ .
 
  Even though initially the Kasner metric has been considered as a cosmological model of the Universe \cite{BP}, we shall use it for the RHIC fireball \cite{SS}. Therefore, the cosmic fluid will be replaced with QGP fluid, having dissipative terms. The corresponding energy - momentum tensor will be written as 
\begin{equation}
T_{\alpha \beta} = \rho u_{\alpha} u_{\beta} + (p - \zeta \Theta) h_{\alpha \beta} - 2 \eta \sigma_{\alpha \beta} ,
\label{5}
\end{equation}
where $\rho$ and $p$ are the mass density and the isotropic pressure, respectively, $u_{\alpha}$ is the comoving 4 - velocity of the fluid , $h_{\alpha \beta} = g_{\alpha \beta} + u_{\alpha} u_{\beta}$ is the projection tensor on a hypersurface orthogonal to $u_{\alpha}$ (the Kasner spacetime being homogeneous, the viscosity coefficients $\eta ~and~ \zeta$  may depend only on time). The scalar expansion is $\Theta = \nabla_{\alpha} u^{\alpha}$ and the traceless shear tensor is given by \cite{HC1} 
\begin{equation}
\sigma_{\alpha \beta} = \frac{1}{2}(h_{\beta}^{\mu} \nabla_{\mu} u_{\alpha}+ h_{\alpha}^{\mu} \nabla_{\mu} u_{\beta})-\frac{1}{3} \Theta h_{\alpha \beta}. 
\label{6}
\end{equation}
Taking the trace of (5) we get
\begin{equation}
T_{\alpha}^{\alpha} = - \rho + 3p - \frac{3 \zeta \lambda}{t}
\label {7}
\end{equation}
where $\lambda \equiv a + b + c $ and the relation $\Theta = \lambda/t$ has been used.

 Einstein's equations (3) yield now \cite{BP,BNOV}
 \begin{equation}
 \lambda - \mu + 12 \pi \zeta \lambda t = 4 \pi (\rho + 3p) t^{2}
 \label{8}
 \end{equation}
 from the $tt$ component and 
 \begin{equation}
 a(1 - \lambda - 16 \pi \eta t) + 4 \pi (\zeta + \frac{4}{3}\eta)  \lambda t = 4 \pi (p - \rho) t^{2}
 \label{9}
 \end{equation}
 from the $xx$ component. For the $yy$ and $zz$ components we must replace $a$ in Eq. (9) with $b~ and~ c$, respectively. In addition, we have $\mu \equiv a^{2} + b^{2} + c^{2}$. 
 
 From the last two equations we see that the following relations should be obeyed
 \begin{equation}
 \rho \propto \frac{1}{t^{2}}, ~~~p \propto \frac{1}{t^{2}},~~~\eta \propto \frac{1}{t},~~~, \zeta \propto \frac{1}{t}.
 \label{10}
 \end{equation}
 The above behaviour of $\rho, p, \eta ~and~ \zeta$ is in accordance with many authors ( see \cite{SS, HC1, BNOV, AB}). \\
 
 \textbf{The isotropic expansion with bulk viscosity}
 
 If we consider $\lambda \neq 1 - 16 \pi \eta t$ in Eq.(9) and its equivalents with $b~ and~ c$ instead of $a$, the space must be isotropic ($a = b = c$). In this case, $\rho~ and~ p$ acquire the form \cite{BP} 
 \begin{equation}
 \rho = \frac{3a^{2}}{8 \pi t^{2}}, ~~~~p = \frac{2a - 3a^{2}}{8 \pi t^{2}} + \frac{3a \zeta}{t} .
 \label{11}
 \end{equation}
 The shear viscosity tensor must vanish now because of the spatial isotropy (the fluid expands isotropically from a pointlike explosion) after a head - on collision of the incoming particles at relativistic velocities). The fluid kinematics would however allow a bulk viscosity term in the stress tensor since the scalar expansion is nonzero. 
 
 Let us now impose $p = - \rho$ as the equation of state of the fluid (this conjecture is justified by the similarities of our system with the state of the Universe at the end of the inflationary period and with the black hole interior, as assumed in \cite{HC1}). We now obtain, from (11)
 \begin{equation}
 \zeta (t) = - \frac{1}{12 \pi t} ,
 \label{12}
 \end{equation}
 an expression already obtained in \cite{HC2}. In addition, Khoury and Parikh \cite{KP} and Parikh and Wilczek \cite{PW} reached also $\zeta < 0$ for the black hole horizon viewed as a membrane. The negative value of the bulk viscosity coefficient was explained by the authors of \cite{PW} through an instability against perturbations, triggering expansion or contraction and reflecting the null hypersurface's natural tendency to expand or contract. Recently, Kolekar and Padmanabhan \cite{KP} obtained a negative $\zeta$ in their study on a thermodynamical extremum principle from which to get the Damour - Navier - Stokes equation where the degrees of freedom varied in the Action Principle are the null vectors in the spacetime instead of the metric tensor. 
 
 The dominant energy condition $|p| \leq \rho$ is obeyed but the strong energy condition $\rho + 3p \geq 0$ is not satisfied, due to the negative pressures. Moreover, Albacete et al. \cite{AKT} and Kovchegov \cite{YK} obtained a negative longitudinal (namely, on the direction of the collisional axis) pressure for the classical gluon dynamics applied to the RHIC at early proper time. We also note that Tawfik \cite{AT} stressed the important role played by the bulk viscosity $\zeta$ in the evolution of the very early universe. This is not surprising if we keep in mind that the hypersurface $t = 0$ is a true singularity for the metric (1), with $a = b = c = 1/2$. We have, indeed
  \begin{equation}
 K = R_{\alpha \beta \mu \nu} R^{\alpha \beta \mu \nu} = \frac{24}{t^{4}}, 
 \label{13}
 \end{equation}
  where the Kretschmann scalar $K$ is divergent at $t = 0$. 
 
 Keepimg in mind that in RHIC the rest masses of the particles are negligible with respect to their kinetic energies, we must have $T_{\alpha}^{\alpha} = 0$. Hence, the Eqs. (7) and (11) yield
 \begin{equation}
 \rho = - p = \frac{3}{32 \pi t^{2}},~~~~\Theta = \frac{3}{2t},~~~~a = \frac{1}{2}.
 \label{14}
 \end{equation}
 It is worthwhile to note that there is no contradiction between the tracelessness of $T_{\mu \nu}$ resulting from conformal invariance and the nonvanishing of the bulk viscosity ( Kubo's formula \cite{KT} ) since our treatment is classical.  In addition, we note that the conservation equation for the energy in the comoving frame \cite{TWMH},
  \begin{equation}
 \dot{\rho} + (p + \rho) \Theta = \zeta \Theta^{2} 
 \label{15}
 \end{equation}
is observed.

For the components of $T_{\mu \nu}$ one obtains
\begin{equation}
T_{t}^{t} = - \rho = - \frac{3}{32 \pi t^{2}},~~~~~T_{x}^{x} = T_{y}^{y} = T_{z}^{z} = \frac{1}{32 \pi t^{2}} = - \frac{p}{3} 
\label{16}
\end{equation}
and the components of $R_{\mu \nu}$ are
\begin{equation}
R_{t}^{t} = - \frac{3}{4 t^{2}}, ~~~~~R_{x}^{x} = R_{y}^{y} = R_{z}^{z} = \frac{1}{4 t^{2}}
\label{17}
\end{equation}
Eq. (7) yields
\begin{equation}
\rho = 3 (p - \zeta \Theta)
\label{18}
\end{equation}
We may introduce an effective pressure $p_{eff} = p - \zeta \Theta$, which is equal to $p$ for a perfect fluid. We see that, in that case, $p_{eff} = \rho/3$. Therefore, if we chose $p_{eff}$ as the physical pressure, the equation of state would become compatible with the tracelessness of $T_{\mu \nu}$, as is required by the conformal invariance at ultrarelativistic energies.

Let us study now the problem of the huge value of the nondimensional entropy per baryon, $\sigma \approx 4. 10^{9}$. As Brevik and Heen \cite{BH} have shown, the above value of $\sigma$ could be obtained by means of the so called ''impulsive'' bulk viscosity in the very early Universe. They concluded that a bulk viscosity during the phase transition near inflationary period (at $t \approx 10^{-33} s$ ) of $\zeta \approx 10^{60} g/cm~ s$ leads to the previous value of $\sigma$. 

With the help of the Eq. (12) for the bulk viscosity in our model for the RHIC isotropic expansion, we reach, using different arguments, the same result as the authors of \cite{BH} : at the moment $t \approx 10^{-33} s$ we obtain $|\zeta| = c^{2}/12 \pi G t \approx 10^{60} g/cm~ s$. One means our model mimics the inflationary stage of the Universe and gives the necessary viscous entropy production , large enough to explain the observed entropy in our Universe.\\

\textbf{Conclusions} 

 The problem of an isotropic dissipative Bjorken expansion is addressed in this paper. Although the pressure of the fluid is minus the energy density, the ''effective'' pressure is one third of the energy density, in accordance with the conformal behaviour of the relativistic particles (the trace of the energy momentum tensor of the relativistic plasma is vanishing). Even though the isotropic nonflat Kasner spacetime is often used for the description of the very early stages of the Universe, we think we have a lot of evidences in the support of its use for RHIC fireball (the deconfined phase). We have also obtained a bulk viscosity for the QGP during its early evolution (near the singularity $t = 0$ ) which is time dependent and, when applied to our Universe, leads to an entropy per baryon very close to the well known value $\sigma \approx 4. 10^{9}$.\\
 
 \textbf{Acknowledgements}\\
 I would like to thank the two anonymous referees whose comments and suggestions helped me to improve the manuscript.


\begin{thebibliography} {23}

\bibitem {SH}
H. Song and U. Heinz, Phys. Lett. B658, 279 - 283 (2008) ( ArXiv : 0709.0742 ).
\bibitem{SO}
S. - Y. Ollitrault, Eur. J. Phys. 29, 275 (2008), ArXiv : 0708.2433 [nucl - th].
\bibitem{MH}
M.P. Heller et al., Acta Phys. Pol. B39, 3183 (2008), ArXiv : 0811.3113 [hep-th].
\bibitem{KK}
K. Kajantie et al., Phys. Rev. D78, 126011 (2008), ArXiv : 0809.4875 [hep-th].
\bibitem{BO}
I. Brevik and S. D. Odintsov, Phys. Rev. D65, 067302 (2002); I. Brevik, Int. J. Mod. Phys. A18, 2145 (2003).
\bibitem{BGG}
I. Brevik, O. Gorbunova and D. Saez-Gomez, Gen. Relat. Grav. 42, 1513 (2010).
\bibitem{TWMH}
A. Tawfik, M. Wahba, H. Mansour and T. Harko, Ann. Phys. (Berlin) 523, 194 (2011).
\bibitem{TH}
A. Tawfik and T. Harko, ArXiv: 1108.5697 [astro-ph].
\bibitem{BENO}
I. Brevik, E. Elizalde, S. Nojiri and S. D. Odintsov, ArXiv: 1107.4642 [hep-th].
\bibitem{MG}
N. Mostafapoor and O. Gron, Astroph. Space Sci. 333, 357 (2011).
\bibitem {BP}
I. Brevik and S. V. Pettersen, Phys. Rev. D56, 3322 (1997).
\bibitem {SS}
S. - J. Sin, S. Nakamura and S. P. Kim, J. High Ener. Phys. 0612, 075 (2006), ArXiv: hep-th/0610113.
\bibitem {HC1}
H. Culetu, Int. J. Mod. Phys. A24, 1593 (2009), ArXiv : hep-th/0701255.
\bibitem {BNOV}
I Brevik, S. Nojiri, S. Odintsov and L. Vanzo, Phys. Rev.D70, 043520 (2004), ArXiv : hep-th/0401073.
\bibitem{AB}
A. Buchel, AIP Conf. Proc. 1031, 196 (2008), ArXiv : 0803.3421 [hep-th].
\bibitem{HC2}
H. Culetu, ArXiv : 0711.0062 [hep-th].
\bibitem{KP}
J. Khoury and M. Parikh, ArXiv : hep-th/0612117 .
\bibitem {PW}
M. Parikh and F. Wilczek, Phys. Rev. D58, 064011 (1998), ArXiv: gr-qc/9712077.
\bibitem{KTP}
S. Kolekar and T. Padmanabhan, ArXiv: 1109.5353 [gr-qc].
\bibitem{AKT}
J. L. Albacete, Y. Kovchegov and A. Taliotis, JHEP 0807: 100.2008, ArXiv: 0805.2927 [hep-th].
\bibitem{YK}
A. Kovchegov, Nucl. Phys. A830, 395c - 402c (2009), ArXiv: 0907.4938 [hep-ph].
\bibitem{AT}
A. Tawfik, Ann. Phys. (Berlin) 523, 423 (2011), ArXiv: 1102.2626 [gr-qc].
\bibitem{KT}
D. Kharzeev and K. Tuchin, ArXiv : 0705.4280 [hep - ph].
\bibitem{BH}
I. Brevik and L. T. Heen, Astrophysics and Space Science 219, 99 - 115 (1994).

\end{thebibliography}
\end{document}